*In vivo* exposure of the bladder using a non-invasive high intensity focused ultrasound toroidal transducer

Victor Delattre[1], Sophie Cambronero[1], Yao Chen[1], Gail ter Haar[2], Ian Rivens[2], Gerry Polton[3], Cyril Lafon[1,2] and David Melodelima[1]

[1]LabTAU, INSERM, Centre Léon Bérard, Université Lyon 1, Univ Lyon, F-69003, LYON, France

[2]Joint Department of Physics, Institute of Cancer Research and Royal Marsden Hospital NHS Trust, Sutton, Surrey, UK

[3] North Downs Specialist Referrals, Bletchingley, Surrey, UK

# ABSTRACT

A toroidal high-intensity focused ultrasound (HIFU) transducer was used to expose normal bladder wall tissues non-invasively *in vivo* in a porcine model in order to investigate the potential to treat bladder tumors. The transducer was divided into 32 concentric rings with equal surface areas, operating at 2.5 MHz. Eight animals were split into two groups of 4. In the first group, post-mortem evaluation was performed immediately after ultrasound exposure. In the second group, animals survived for up to seven days before post-mortem evaluation. The ultrasound imaging guided HIFU device was hand-held during the procedure using optical tracking to ensure correct targeting. One thermal lesion in each animal was created using a 40 s exposure at 80 acoustic Watts (free-field) in the trigone region of the bladder wall. The average (±Standard Deviation) abdominal wall and bladder wall thicknesses were $10.3 \pm 1.4$ mm and $1.1 \pm 0.4$ mm respectively. The longest and shortest axes of the HIFU ablations were $7.7 \pm 2.9$ mm and $6.0 \pm 1.8$ mm, respectively, resulting in an ablation of the whole thickness of the bladder wall in most cases. Ablation were performed at an average depth (distance from the skin surface to the centre of the HIFU lesion) of $42.5 \pm 3.8$ mm and extended throughout the thickness of the bladder. There were two cases of injury to tissues immediately adjacent to the bladder wall but without signs of perforation, as confirmed by histological analysis. Non-invasive HIFU ablation using a hand-held toroidal transducer was successfully performed to destroy regions of the bladder wall *in vivo*.

# INTRODUCTION

Bladder cancer is the sixth most common type of cancer. Muscle-invasive bladder cancer is associated with poor prognosis with a 5-year overall survival of only 35% (Ferlay et al., 2018). Patients are offered radical cystectomy and/or chemotherapy and/or radiotherapy (Tan et al., 2023). Successful treatment can result in impaired quality of life (Taarnhøj et al., 2019). There is a significant need for better treatments for these patients. High Intensity Focused Ultrasound (HIFU) may be used as a safe, non-invasive, modality for the treatment of bladder tumors (Watkin et al., 1996). Despite this method's clinical potential, there are still challenges to overcome. Before reaching their target in the body, the ultrasound waves must propagate through skin, fat and muscle layers. These tissues have different acoustic properties, (including acoustic impedance, speed of sound and attenuation), that alter the focusing effect and energy deposition at the focus (Cambronero et al., 2023; Jung et al., 2011). Skin or subcutaneous burns, usually mild but occasionally severe (grade 3, requiring surgical intervention) have frequently (15%) been reported in clinical studies (Sehmbi et al., 2021) and represent the most common side effect of HIFU treatments. Another limitation of HIFU exposures is that the ablations are typically cigar-shaped, with millimetric dimensions (Haar & Coussios, 2007). As a consequence, the clinical treatment of solid tumors requires mechanical motion of the transducer to juxtapose many of single non-spherical ablations until the entire tumor volume is covered. Such a procedure involves long treatment times which have been mitigated by implementation of an elaborate robotized arm combined with complex electronics (Napoli et al., 2020).

We have reported that at the early clinical stage, the use of a toroidal HIFU transducer shows promise for treating liver metastases. This toroidal HIFU transducer achieved fast, selective, safe and well-tolerated large volume tissue ablation up to 10 cc in one minute of treatment (Battais et al., 2020) allowing to use the device by hand (N'Djin et al., 2015; Vincenot et al., 2013). To date, 30 patients have been included in a phase I-II clinical trial of this device (Dupre et al., 2019; Dupré et al., 2023). Thanks to this initial experience, the first non-invasive use of such a transducer has now been reported for lesioning the liver (Cambronero et al., 2023) and the placenta (Caloone et al., 2017). These features might overcome the current technical limitations of HIFU by enabling the treatment of the entire tumor volume in a single HIFU exposure without the need for mechanical displacement of the device. In the present study the use of this device for treating the bladder was investigated in a large animal model before progressing to clinical trials in dogs and then in humans.

# MATERIALS AND METHODS

The toroidal HIFU device, which has already been described previously (Cilleros et al., 2021) has an active diameter of 67 mm and a radius of curvature of 70 mm. The transducer was divided into 32 concentric ring-shaped emitters with identical surface areas (78 mm$^2$), all operating at 2.5 MHz. The average real and imaginary parts of the acoustic impedance for the 32 emitters were 58.4 ± 1.7 ohms (52.9 to 61.9) and -7.0 ± 1.6 ohms (-3.6 to 10.0) respectively. The air-backed HIFU transducer was a piezocomposite material (Imasonic, Voray-sur-l'Ognon, France). Because of the toroidal geometry, the focal zone is a ring 30 mm in diameter, at a depth of 70 mm. Moreover, because of the geometrical characteristics of a torus, the ultrasound beams coming from each of the 32 emitters intersect between the transducer and the focal ring to form a secondary focal zone (Figure 2). The pressure in this additional focal zone is higher than the pressure in the focal ring and can therefore be used to create large ablations (Melodelima, et al. 2006). An example of the pressure field is shown in (Cambronero et al., 2023). A sectorial ultrasound imaging probe working at a frequency of 7.5 MHz (Vermon, Tours, France) was placed at the center of the HIFU transducer and connected to an ultrasound scanner (EB4012, B-K Medical, Herlev, Denmark) to guide the exposure. The ultrasound imaging plane was aligned with the HIFU acoustic axis. Both transducers were placed in acoustic contact with the skin using coupling water (Baxter, Deerfield, IL, USA) contained in a polyurethane cover (CIV-Flex Transducer cover, CIVCO, Kalona, IA, USA). This coating attenuated the ultrasound pressure by approximately 2% at 2.5 MHz. To prevent the transducer from becoming too hot during the HIFU procedure, the water was cooled at 6°C and flowed at a continuous rate of 0.6 L/min using a peristatic pump (Masterflex L/S Model 7555-05, Cole_Parmer Instruments Co., Chicago, IL, USA) in a closed cooling circuit. Additional ultrasound imaging was performed using an Aplio a450 version 2 ultrasound scanner and a 12-MHz abdominal 8C1 probe, (Canon Medical Systems, Otawara, Japan). This was used to measure abdominal wall thickness and to localize the trigone region, measuring its distance to the skin surface and precisely target the trigone. The trigone region is located on the posterior bladder wall, between the ureter insertion and the urethra (Fig. 1). It was chosen as it can be consistently identified between animals. It also provides a natural target for safety assessment as it is close to the ureters and to the rectum. Moreover, tumors located near the trigon are extremely difficult to surgically resect. Two specific holders with optical markers were designed to fit to the therapeutic probe and the 12-MHz abdominal ultrasound imaging probe. The position of the 12-MHz abdominal 8C1 imaging probe was recorded using an optical tracking camera (Polaris Vicra, NDI, Waterloo, Ontario, CA). Preliminary calibration was done to ensure that the position of the therapeutic probe can be aligned with the one of the 12-MHz abdominal ultrasound imaging probe with an accuracy of less than 1 mm.

The HIFU transducer drive was similar to that previously reported (Chenot et al., 2010). The custom-made amplifier (Image Guided Therapy, Pessac, France) has 32 channels programmable in phase and power over a frequency bandwidth of 1 to 5 MHz. The phase of each channel could be adjusted with a resolution of 1 degree. The electrical power delivered by each channel could be adjusted up to 20 W.

The time needed to change the phase and/or power delivered was 1 ms. The spatial repartition and intensity of the pressure field produced can be controlled electronically by modulating the amplitude and the phase applied to each of the 32 individual transducers. The phase, emitted power and reflected power of each channel were measured and recorded during the HIFU sonications using a custom-made user interface. The positioning of the HIFU device was made according to the previously recorded position of the 12-MHz abdominal 8C1 probe. The user interface displayed the position of the HIFU focal region superimposed on the sonogram, making it possible to target the HIFU ablation in the tissues and to try to visualize the treated zone immediately after each sonication.

*Animal preparation and anaesthesia*

All procedures described in this article were carried out in the Laboratory of Experimental Surgery (Institut de Chirurgie Experimentale, Centre Léon Bérard, Lyon, France: DDPP D693880501). The animal experiments were performed under a research protocol approved by the local approved ethical committee n°10 of animal experimentation and authorized under the number APAFIS#32724-2021061612029203v2. These experiments were conducted in accordance with European legislation covering the use of animals for scientific purposes (Directive 2010/63/EU). Fine tuning of the experimental parameters and procedure was achieved using fresh pig cadavers from other experimental protocols, in order to minimise the number of live animals used for this study.

The experiments were conducted on eight female Landrace pigs from a registered supplier (EARL Porc du plateau, Saint Clair sur Galaure, France), 8-9 weeks old, weighing between 14 and 18 kg. The porcine model was specifically chosen because its anatomy and physiology are very close to those of humans. The animals were housed in social groups with balls, teething toys and a basket. The animals were acclimatised on-site 7 days before the start of the experiments and were fasted 24 h before the HIFU procedure. Premedication was achieved using an intramuscular injection of a mix of ketamine (15 mg/kg), azaperone (2.2 mg/kg) and atropine (0.5 mg) 15 min before anaesthesia. A 20-gauge catheter was then placed in an auricular vein to induce anaesthesia with 0.4 ml/kg of KZX which is a mix of 2.5 ml Rompun® 2% and 2.5 ml Imalgène 1000® in a flask of Zoletil® 100 powder. Anaesthesia was then maintained with 0.2ml/kg of KZX every 30 min. Two pigs were injected with diuretic (1mL furosemide) because their bladder did not appear full on ultrasound images. Hydration was provided by isotonic perfusion of a physiological salt solution at 0.9 %. During the HIFU procedure, $SpO_2$ and temperature were monitored. The animals were positioned in a dorsal decubitus position on an air heated mattress. The skin at the site of the HIFU exposure was washed and shaved if needed.

Animals were visited at least twice daily. Food consumption and behaviour were scored as follows: food intake (0: ration eaten in less than one hour, 1: ration eaten in more than one hour, 2: majority of ration eaten, 3: anorexia); behaviour (0: normal interactions and tonus, 1: weak tonus but presence of activity,

2: decubitus but reaction to stimulation, 3: decubitus without reaction). If the total score is 4 or more, a subcutaneous injection of methylprednisolone (10mg/kg) is administrated and if no amelioration is observed within 6 hours, the animal is euthanized. Additionally, after HIFU exposure, presence of blood in the faeces and/or urine, absence of urine, pain during micturition are humane endpoints because they can be consequences of bladder or rectal rupture.

*HIFU procedure*

A 12-MHz ultrasound imaging probe was used to image the region of the bladder to be targeted (trigone) and to measure the thickness of all tissues between the probe and the focus, including the skin, fat and muscle layers. The HIFU probe was then placed on the skin of the animal such that the imaging plane of the ultrasound imaging probe integrated with the HIFU transducer corresponded to the image obtained with the 12-MHz ultrasound imaging probe. The HIFU device was hand-held. A thin layer of demineralized water was placed between the HIFU device and the skin to ensure perfect acoustic coupling. Acoustic coupling was verified using the image provided by the integrated ultrasound imaging probe at the center of the HIFU transducer. In each animal the trigone was targeted in one location only. Prior to animal experiments, the attenuation coefficients of the porcine abdominal wall layers (i.e., skin, fat and muscle) were measured and numerical simulations were performed to define exposure parameters (acoustic power, time of exposure, focal shape and location) as described by (Cambronero et al., 2023). Since the abdominal wall thickness and the depth of the trigone beneath the skin were consistent in all animals, all HIFU exposures were performed with the same parameters. The quantity of cooling water contained in the polyurethane cover was adjusted according to the filling of the bladder to keep the target (trigone) in the focal zone. All exposures consisted of two consecutive sonications of 20 s each. The maximum pressure zone was first located 36 mm below the skin creating a -6 dB toroidal focus of 6 mm in width and 3 mm in height. Without moving the HIFU device and with no delay, a second sonication was performed by electronically setting the depth of the maximum pressure zone to be 34 mm deep below the skin creating a -6 dB toroidal focus of 2 mm in width and 3 mm in height. This slight change of the focal location was made to minimize overheating and boiling. The free field acoustic power ($P_a$) was set to 80 W corresponding to an energy of 3.2 kJ. $P_a$ was calculated from the measured electrical power during each sonication and the average efficiency of the transducer (60%). The efficiency of the transducer was measured using an acoustic balance as described by (Davidson, 1991). The derated value of the acoustic power ($P_d$) was then calculated using the following equation by considering the attenuation in the pre-focal tissues and a total transmission coefficient $T = 0.96$ as described in a previous study (Cambronero et al., 2023):

$$P_d = P_a \cdot T \cdot 10^{-\frac{\alpha(z_s, z_f, z_m, z_b)}{10\, dB}} \qquad \text{Equation 1}$$

The attenuation in the pre-focal tissues was computed using Equation 2:

$$\alpha(z_s, z_f, z_m, z_b) = \alpha_s z_s + \alpha_f z_f + \alpha_m z_m + \alpha_b z_b \qquad \text{Equation 2}$$

where $\alpha_s$=3.9 dB.cm$^{-1}$, $\alpha_f$=1.1 dB.cm$^{-1}$ and $\alpha_m$=2.0 dB.cm$^{-1}$ are the measured attenuations in pig tissues (dB.cm-1) of the skin, fat and muscle respectively at 2.5MHz. The attenuation value used for the bladder ($\alpha_b$) was 1.2 dB.cm$^{-1}$ (Hasgall et al., 2011) at 2.5MHz. $z_s$, $z_f$, and $z_m$ are the thicknesses (cm) of the skin, fat and muscle respectively, and finally $z_b$ is the thickness of the bladder wall measured before each exposure (cm).

When the exposure was completed, the 12-MHz ultrasound imaging probe was used to image the bladder to look for echogenicity changes associated with the lesions. The skin was examined for any macroscopic signs of damage (erythema or skin burns).

The first 4 animals were used to assess acute damage, and were euthanized within 10 min of exposure and ultrasound imaging, using an intravenous injection of 0.16mL/kg of Euthoxin® (72.9 g/kg pentobarbital). A laparotomy was then performed to examine the bladder and adjacent organs for any macroscopic signs of damage. The treated area was measured along two perpendicular axes.

A subsequent survival study to establish the longer-term risk of bladder perforation and injury to adjacent pelvic organs involved 4 animals. After exposure, the animals were recovered from anaesthesia and observed for signs of anorexia, distress, difficulty in micturition or haematuria, fever and changes in general behaviour. Seven days after HIFU exposure, the animals were premedicated as described before. The bladder was scanned with ultrasound, before euthanasia with Euthoxin®. The abdomen was opened in order to remove the bladder, and to look for potential damage in adjacent organs. The treated area was measured on the outside surface of the bladder along two perpendicular axes. The samples were inspected visually for macroscopic change (see Data Analysis) and representative samples were fixed in a 4% solution of formaldehyde. After 48 h, the samples were transferred to phosphate-buffered saline, dehydrated using increasing concentrations of alcohol, treated with intermediate medium and embedded in paraffin. The embedded samples were sliced (3 µm thick) and stained using hematoxylin and eosin. The tissues were examined with brightfield microscopy.

*Data analysis*

A photograph of each excised bladder sample was taken. Images of the lesions were then analysed using the Measure tool of ImageJ, version 1.52a (National Institutes of Health, Bethesda, MD, USA; http://imagej.nih.gov/ij). The longest and shortest axes of the treated area visible as a white necrotic core on the tissue sample surface were measured. All measured data are given as mean values ± standard deviations (minimum value to maximum value).

RESULTS

*Overview of response to HIFU exposures*

Food consumption and behaviour scores were 0 for all animals during the 7 days prior to exposure. All animals tolerated the exposure well over the experimental period. The animals followed for 7 days after exposure recovered from anaesthesia within 2 h of the procedure end and quickly resumed eating and normal behaviour with no signs of pain or fever. Animals showed no signs of distress, difficulty in urination or had haematuria during the observation period. No humane endpoints were reached. During the seven days of observation all scores remained at 0. Table 1 provides an overview of the exposure parameters for each of the 4 non-recovery and 4 recovery animals. Seven of the 8 HIFU exposures resulted in visible damage at the target site on post-mortem examination. There was no evidence of bladder perforation. Post-mortem examination of the acute exposures revealed clear evidence of bladder lesioning (Figure 3) except in one case (A4) where the animal moved during exposure. In this specific case, technical issues with the experimental set up increased the duration of the experiment and anaesthesia. Acute HIFU lesions appeared white and homogeneous and were all located at the trigone location (Figure 3a). In pig A2, injury of the rectal wall and the presacral nerve was observed but no histological analysis was performed on this animal. Similarly, for the animals followed up to seven days, all exposures resulted in a HIFU lesion in the bladder wall at the trigone location (Figure 3b) as seen with gross pathological analysis. The abdominal wall thickness for all animals was on average $10.3 \pm 1.4$ mm (8.8 to 13.3). According to the filling of the bladder, the target region (trigone) was located at an average depth of $40.5 \pm 3.8$ mm (36.0 to 48.5). Therefore, the quantity of cooling water contained in the polyurethane cover was adjusted such as the distance between the transducer and the skin was on average $18.6 \pm 4.0$ mm (13.0 to 23.0). Adjusting the quantity of water allowed to compensate for the variations of the target distance. Since this variation was due to the quantity of urine in the bladder it was not needed to change the acoustic power assuming that the attenuation of urine is close to the one of water. There was no visible damage to the skin and intervening tissues. HIFU lesions were not visible as a hyperechoic region in the bladder wall on ultrasound images immediately after the sonication. The longest and shortest axes of the HIFU lesions on the surface of the bladder were $7.7 \pm 2.9$ mm (5.0 to 13.0) and $6.0 \pm 1.8$ mm (5.0 - 10.0), respectively.

Coagulation necrosis was confirmed by histological analysis except for B1 for which coagulation necrosis was not confirmed. When present, coagulation necrosis extended through the entire thickness of the bladder wall. A typical example of the histological analysis is shown in Figure 4 corresponding to HIFU ablation in animal B4. Histological analysis showed focal necrosis of the serous membrane and of the mucosal membrane, indicating a clear lesion due to thermal ablation. There was injury to tissues immediately adjacent to the bladder wall in two animals followed for seven days after exposure (B3 and B4). In these two cases the rectal wall was inflamed but there were no signs of perforation seen histologically.

## DISCUSSION

This study has shown that non-invasive ultrasound imaging guided transabdominal HIFU exposure performed using a hand-held toroidal transducer can successfully ablate the porcine bladder wall trigone. Exposures allowed the safe creation of thermal ablations measuring, on average, 7.7 mm by 6.0 mm and extending through the entire thickness of the bladder wall. The HIFU exposures lasted 40 s. All exposures were well tolerated and, after recovery from general anaesthetic, did not affect the animals' behaviour and food consumption (all scores remained at pre-exposure baseline levels).

The exposure parameters were calculated according to the methods already presented (Cambronero et al., 2023) in which the thickness and attenuation of intermediate tissues is used to determine the exposure *in situ* acoustic power. This allowed the bladder wall, which is about 2-4 mm thick, to be exposed precisely without major clinical adverse effects. Monitoring tissue damage during and after exposure with conventional B-Mode ultrasound imaging would be a valuable aid in determining the therapeutic efficacy but was not possible in this work. This could be attributed to the thin bladder wall, making it challenging to clearly observe any echogenic changes after HIFU exposure. However, since a single exposure is anticipated to ablate the entire tumor, this may not be an absolute limitation. Clinical results in tumors may also vary, as seen in previous reports for liver treatments with a more distinct evaluation of the treated area (Dupré et al., 2015). Nevertheless, other clinical methods may need to be used to monitor cancer treatment. Magnetic resonance imaging (MRI) allows use of contrast enhanced MR for post-treatment monitoring and also use of MR thermometry during exposure with typical spatial resolution in the order of 1 mm and a temperature resolution less than 1°C (Siedek et al., 2019). The MRI compatibility of the device has not been tested, although all materials and manufacturing processes have been designed in accordance with previous research (Melodelima et al., 2006)for potential MRI applications. Many ultrasound-based methods have also been described to estimate the HIFU-treated regions *in vivo* such as elastography (Barrere et al., 2020), quantitative ultrasound (Ghoshal et al., 2016), contrast-enhanced ultrasound (Apfelbeck et al., 2018) or ultrasound thermometry (Lewis et al., 2015). Ultrasound imaging has higher spatial resolution (typically in the order of 100 µm) than MRI. However, ultrasound imaging does not provide reliable thermometry in the range of temperature reached during HIFU exposure (typically more than 70°C).

The overall morbidity in this study was low. There was no evidence of urinary extravasation. No animal experienced voiding difficulties even though the exposure targeted the bladder trigone. Inflammation of the rectal wall in 2/8 animals was not unexpected as the focus was placed at the posterior bladder wall surface when targeting the trigone which is in contact with the rectal wall, and the device is designed to treat tumors of up to 3 cm in diameter rather than a thin (2-4 mm) membrane. Seven days post-exposure, there was no evidence of rectal fistula and histological analysis confirmed the macroscopically observed inflammation but showed an absence of necrosis. Importantly, damage to the skin or subcutaneous fat was not observed in this study.

The pre-clinical study described in this article was performed on pigs because of their similar physiology to humans. There is no established bladder tumor model available in pigs or any other large animal in which to evaluate HIFU treatments. We believe that this limitation is the main reason for the observed complications. However, it is also important to note that during clinical trials, the tumors will be in contact with the rectal wall. Therefore, even if focusing delays and/or acoustic power could have been reconsidered, it is particularly comforting that this study demonstrates the safety of the treatment when HIFU exposures are carried out on the rectum without causing fistulas. The toroidal HIFU transducer, which has demonstrated its ability in other tissues (Caloone et al., 2017; Dupré et al., 2023; N'Djin et al., 2011) to create relatively large lesions (up to 50 cm$^3$) over relatively short time periods (370 s) that are visible on sonograms, may represent a powerful technique for treating bladder tumors non-invasively. The probe can be hand-held, making it easy to use. Moreover, previous studies that used the same transducer demonstrated that using electronic steering the focal depth could be adjusted from 10 mm to 50 mm below from the skin (Sanchez et al., 2021). This can be useful for treating different size bladder tumors in a variety of locations (Tang et al., 2021). In the absence of a large animal model of bladder cancer, the adequacy of the exposure parameters and spacing of potentially juxtaposed lesions can only be confirmed in bladder tumors. This can be done with human or dog patients that are prone to bladder tumors (de Brot et al., 2018). Spontaneous canine tumors have several features that make them a good surrogate for testing experimental treatments before their introduction into the hospital clinic. They are similar to those that develop in humans, with respect to their histological and genetic features, and metastatic behaviour (Fulkerson et al., 2017; Knapp et al., 2000). In addition to providing a good model for bladder tumors, validation of the treatment on dogs could lead to the use of HIFU in this veterinary application. Current treatment options for canine trigonal tumors comprise palliative medical approaches with limited efficacy or radical surgical approaches with very high morbidity (Fulkerson & Knapp, 2015). The porcine model used in this study is close in size to the dogs that most frequently develop bladder tumors (Mutsaers et al., 2003) and therefore, this treatment could be translated to the veterinary clinic with minimal adaptation. There is no database of dog bladder sizes and thicknesses as they change proportional to patient size but the young pigs were selected for their overall comparability with small terrier breed dogs. Moreover, it is possible to modify bladder volume by filling it with saline. Bladder wall thickness at any given time is largely defined by degree of filling. In addition, the distance from skin to serosa can also be adjusted through that mean. From the point of view of comparability with humans: in our study the skin to serosa distance measured on the young pigs with a partially filled bladder ranges from 36.0 mm to 48.5mm as mentioned in table 1. The anterior bladder wall of a full bladder in adults is located approximately between 40 and 80 mm (Oelke et al., 2006). By adjusting the filling of the bladder and the focusing of the probe we are confident in the possibility to target bladder wall in humans.

In conclusion, this study has demonstrated that it is feasible to use a toroidal transducer to create large ablations in short time (40 s) at predetermined sites in the bladder wall, non-invasively.

ACKNOWLEDGEMENTS

The authors thank the staff of the Laboratory for Experimental Surgery for their assistance with the animal study. This work was funded by the FUS foundation.

DECLARATION OF INTERESTS

The authors declare that they have no known competing financial interests or personal relationships that could have appeared to influence the work reported in this paper.

FIGURES

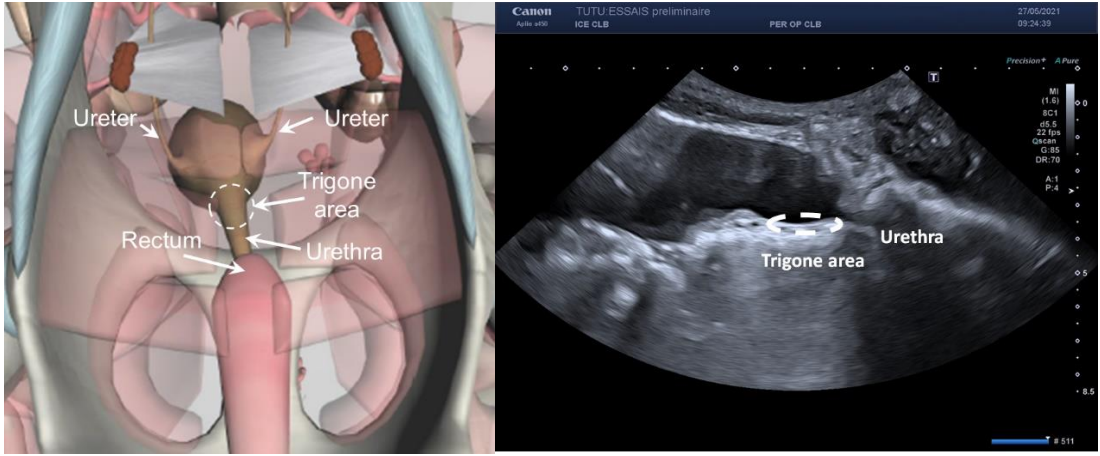

Figure 1. (a) 3D anterior view of the bladder area in a pig. (b) Ultrasound image of the bladder and urethra in sagittal view.

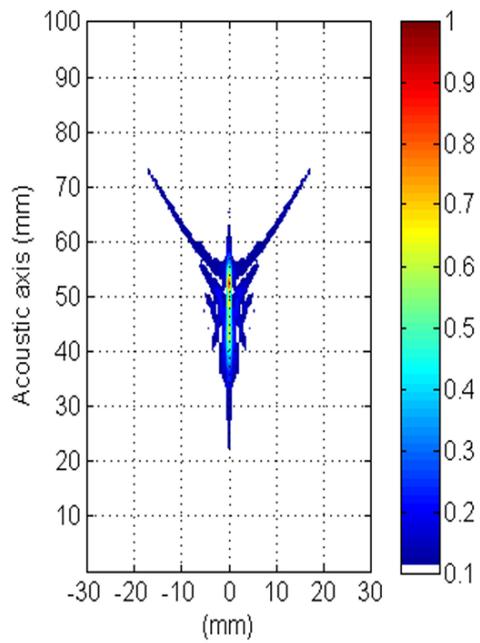

Figure 2. Pressure field in water of the natural focusing of the toroidal transducer.

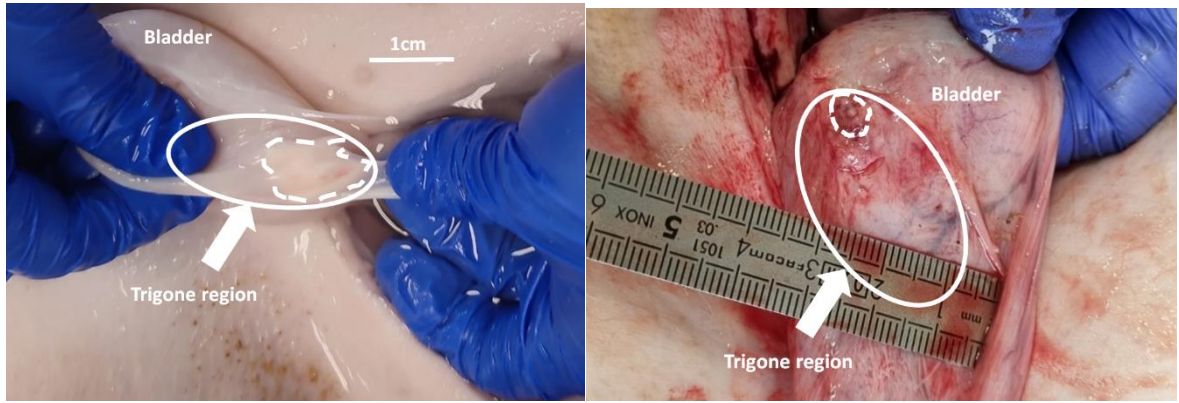

Figure 3. The macroscopic appearance of HIFU lesions created in the bladder wall. (a) Acute treatment. (b) Seven days after treatment, lesion is circled with a white dotted line and arrows point towards the trigone region.

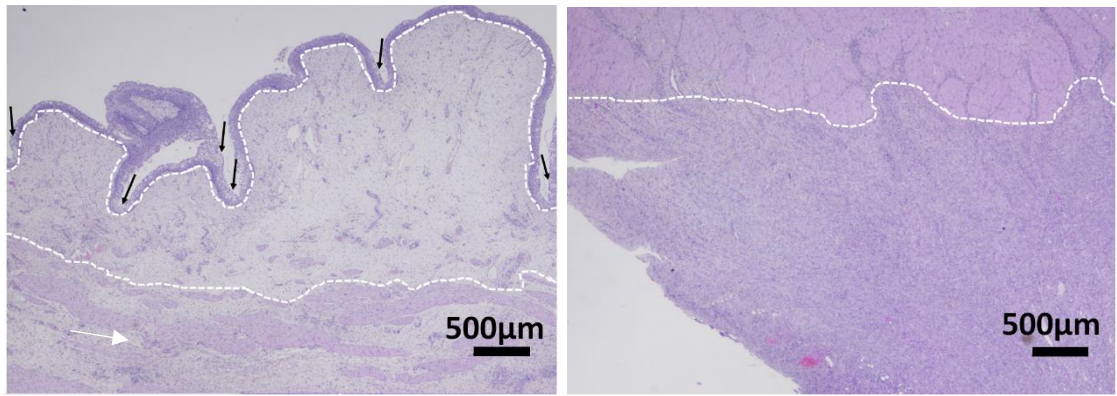

Figure 4. Histological results. Microscopic view of the bladder wall tissue after staining with H&E. (a) Treated mucosal membrane of the bladder. Black arrows show focal degeneration of the epithelium and oedema is mostly located between the white dotted lines. (b) The transition between treated and untreated (delineated by the white dotted line) tissue is a few micrometers. Fibrovascular proliferation was observed in the treated zone which is below the white dotted line.

TABLE

| Animal # | Distance from skin to focus (mm) | $P_a$ (W) | $P_d$ (W) | Lesion long axis (mm) | Lesion short axis (mm) | Comments |
|---|---|---|---|---|---|---|
| A1 | 43,5 | 80 | 55 | 10 | 10 | |
| A2 | 45,6 | 76 | 31 | 10 | 5 | |
| A3 | 48,5 | 75 | 44 | 8 | 5 | |
| A4 | 40,2 | 81 | 44 | N/A | N/A | No lesion |
| B1 | 43,0 | 78 | 50 | 6 | 6 | Coagulation necrosis not confirmed |
| B2 | 40,2 | 80 | 47 | 7 | 5 | |
| B3 | 36,0 | 82 | 42 | 6 | 5 | Inflamed rectal wall but no perforation |
| B4 | 42,6 | 82 | 47 | 5 | 5 | Inflamed rectal wall but no perforation |

Table 1. Exposure parameter overview and lesions dimensions observed on gross pathology. Animals A1-A4 were killed within one hour of exposure. Animals B1-B4 were recovered and survived for 7 days. The applied acoustic power was computed from the measurement of the electrical power during sonication, considering the efficiency of the transducer (60%). The derated acoustic power was computed from eqn (1).